\documentstyle[preprint,eqsecnum,aps]{revtex}
\newcommand{\be}{\begin{equation}}
\newcommand{\ee}{\end{equation}}

\newcommand{\qij}{\mbox{$q_{i,j}$}}
\def\simlt{\mathrel{\lower .3ex \rlap{$\sim$}\raise .5ex \hbox{$<$}}}
\def\simgt{\mathrel{\lower .3ex \rlap{$\sim$}\raise .5ex \hbox{$>$}}}
\def\q{$ {\bf q}$}

\begin{document}
\draft
\tighten
\title{A Model for Force Fluctuations in Bead Packs}
\vspace{1cm}

\author{S.N. Coppersmith\thanks{AT\&T Bell Laboratories,
Murray Hill, NJ, 07974}\thanks{The James Franck Institute,
The University of Chicago, 5640 Ellis Avenue,
Chicago, IL  60637 (present address)}
\and
C.-h. Liu\thanks{Exxon Research \& Engineering Company,
Route 22 East, Annandale, NJ  08801}
\and
S.~Majumdar\thanks{Department of Physics, Yale University,
New Haven, CT 06511}
\and
O.~Narayan\thanks{Department of Physics, Harvard University,
Cambridge, MA  02138 }\thanks{Present address:
Department of Physics, University of California, Santa Cruz,
CA  95064}
\and
T.A.~Witten\footnotemark[2]
}
\date{\today}

\maketitle

\baselineskip = 20pt

\begin{abstract}
We study theoretically the complex network of forces
that is responsible for the static structure and
properties of granular materials.
We present detailed calculations for a model in which the
fluctuations in the force distribution arise because of
variations in the contact angles and the constraints
imposed by the force balance on each bead of the pile.
We compare our results for force distribution function
for this model, including exact results
for certain contact angle probability distributions, with
numerical simulations of force distributions
in random sphere packings.
This model reproduces many aspects of the
force distribution observed both in
experiment and in numerical simulations of sphere packings.

Our model is closely related to some that have been studied
in the context of self-organized criticality.
We present evidence that in the force distribution context,
``critical'' power-law force distributions occur only when
a parameter (hidden in other interpretations) is tuned.
Our numerical, mean field, and exact results all indicate
that for almost all contact distributions the
distribution of forces decays exponentially at large forces.
\end{abstract}



\baselineskip = 20pt

\newpage
\section{Introduction}
Disordered geometric packings of granular materials\cite{sandrefs}
have fascinated researchers for many years.\cite{bernal}
Such studies, with their applicability to the geometry of
glass-forming systems, initially were concerned with categorizing
the void shapes and densities.
More recently, partly in recognition of the ubiquity of granular
materials and their importance to a wide variety of technological
processes, interest has focused on how the forces
supporting the grains are distributed.
Visualizations of two-dimensional granular
systems\cite{2dbirefringence} demonstrate weight concentration
into ``force chains.''
It is natural to expect that similar concentrations of
forces will occur in three dimensions.
The distinctive forces in bead packs also give rise to distinctive
boundary-layer flow\cite{flowrefs} and novel sound-propagation
properties.\cite{sound.in.sand.paper}

Ref.~\cite{liu95} presents experiments, simulations, and
theory characterizing the inhomogeneous forces
that occur in stationary three-dimensional bead packs,
focussing particularly on the relative abundance of forces that
are much larger than the average.
If the bead pack were a perfect lattice, then, at any given depth,
no forces would be greater than some
definite multiple of the average force.
At the other extreme, if the network of force-bearing contacts
were fractal,\cite{fractal} then
fluctuations in the forces (characterized, say, by their
variance) would become arbitrarily large compared to the
average force at a given depth, as the system size is increased.
Ref.~\cite{liu95} shows that the forces in bead packs
are intermediate between these two extremes.
The forces are unbounded, but
the number of large forces falls off exponentially
with the force.
The fluctuations remain roughly the same as the
average force, regardless of how large the bead pack becomes.
A simple model was introduced to understand the
results of the experiments and simulations.

This paper presents the detailed analysis of the
model introduced in Ref.~\cite{liu95}.
The model yields force distributions which agree
quantitatively with those obtained in
numerical simulations of sphere packings.
Generic distributions of contacts lead to force distributions
which decay exponentially at large forces, though a special
distribution exists for which the force distribution
is power law.
We discuss the relationship of this model to other related
systems as well as present the analysis leading to the
results that are quoted in Ref.~\cite{liu95} without derivation.

The paper is organized as follows.  Section~\ref{sec:model}
defines the model, discusses several limiting cases that have
been discussed previously in other contexts, and then presents
our analysis of the force distribution expected in the
context of force chains in bead packs.
Special emphasis is placed on one particular contact distribution,
the ``uniform'' distribution, which is the most random
distribution consistent with the constraint of force balance.
We first present a mean field solution for this model, and
then show that this mean field solution is exact.
We also obtain exact results for a countable set of
non-generic distributions as well as mean-field and numerical
results for other contact distributions.
Evidence is presented that almost all contact distributions lead
to exponentially decaying force distributions.
Section~\ref{sec:simulation} discusses numerical simulations
of sphere packings, which we analyze to obtain contact
probability distributions to be used in the $q$-model.
We show that the force distribution predicted by the
model with this contact distribution agrees quantitatively
with the force distribution in the simulation.
Appendix A presents some mathematical identities concerning
the uniform $q$-distribution which are used in the text.


\section{The \q-model}
\label{sec:model}
\subsection{Definition of the model}
\label{subsec:modeldef}

Here we introduce the model, which assumes
that the dominant physical mechanism leading
to force chains is the inhomogeneity of
the packing causing an unequal distribution of
the weights on the beads supporting a given grain.
Spatial correlations in these fractions as well as
variations in the coordination numbers of the grains are ignored.
We consider a regular lattice of sites, each with
a particle of mass unity.
Each site $i$ in layer $D$ is connected
to exactly $N$ sites $j$ in layer $D+1$.
Only the vertical components of the forces are considered
explicitly (it is assumed that the effects of the
horizontal forces can be absorbed in the random variables
$q_{ij}$ defined below).
A fraction $q_{ij}$ of the total weight supported by
particle $i$ in layer $D$ is transmitted to particle $j$
in layer $D+1$.
Thus, the weight supported by the particle
in layer $D$ at the $i^{th}$ site,
$w(D, i)$, satisfies the stochastic equation:
\be
w(D+1, j) = 1 + \sum_i \qij (D) w(D, i)\ \ .
\label{modeleqn}
\ee
We take the fractions $q_{ij}(D)$ to be random variables,
independent except for the constraint $\sum_j q_{ij} = 1$,
which enforces the condition of force balance on each particle.
We assume that
the probability of realizing a given assortment of
$q$'s at each site $i$ is given by a distribution function
$\rho(q_{i1},\ldots,q_{iN}) = \{\prod_jf(q_{ij}) \}
\delta(\sum_j q_{ij} - 1)$.
We define the induced distribution $\eta(q)$ as:
\be
\eta(q) = \prod_{j \ne k}\int {dq_{ij}}
\rho(q_{i1},\ldots,q_{ik}=q,\ldots,q_{iN})\ .
\label{eq:etadef}
\ee
Because $\rho(q_{i1},\ldots,q_{iN})$ is a probability
distribution and $\sum_{i=j}^N q_{ij} = 1$,
the induced distribution must satisfy the conditions
$\int_0^1 dq \eta(q) = 1$, $\int_0^1 dq \ q\ \eta(q) = 1/N$.

In this paper
we focus on the force distribution $Q_D(w)$,
which is the probability that a site at depth $D$
is subject to vertical force $w$.
We obtain $Q_D(w)$ for different distributions of $q$'s.
We will also consider the force distribution $P_D(v)$
for the normalized weight variable $v=w/D$.
For $\eta(q) = \delta(q-1/N)$, where each particle distributes
the vertical force acting on it equally among all its neighbors,
the force distribution at a given depth is homogeneous:
$Q_D(w) = \delta(w-D)$, or $P_D(v) = \delta(v-1)$.
At the other extreme, there is a ``critical'' limit,
when $q$ can only take on the values $1$ or $0$, so
that weight is transmitted to a single underlying particle.
For this, as discussed in the next section, the force distribution
obeys a scaling form and decays
as a power law at large forces, $Q(w) \propto w^{-c}$,
where $c (N \geq 3) = 3/2$ and $c (N=2) = 4/3$.
We demonstrate that this power law does not occur when $q$
can take on the values other than $1$ and $0$, as is the case
for real packings.
Generic continuous distributions of $q$'s lead to
a distribution of weights that, normalized to the mean,
is independent of depth at large D and which decays
exponentially at large weights.
We solve the model exactly for a countable infinite set of
$q$-distributions, and present mean-field and numerical results
for other distributions of $q$'s.

\subsection{The \q-model for the ``critical'' case}

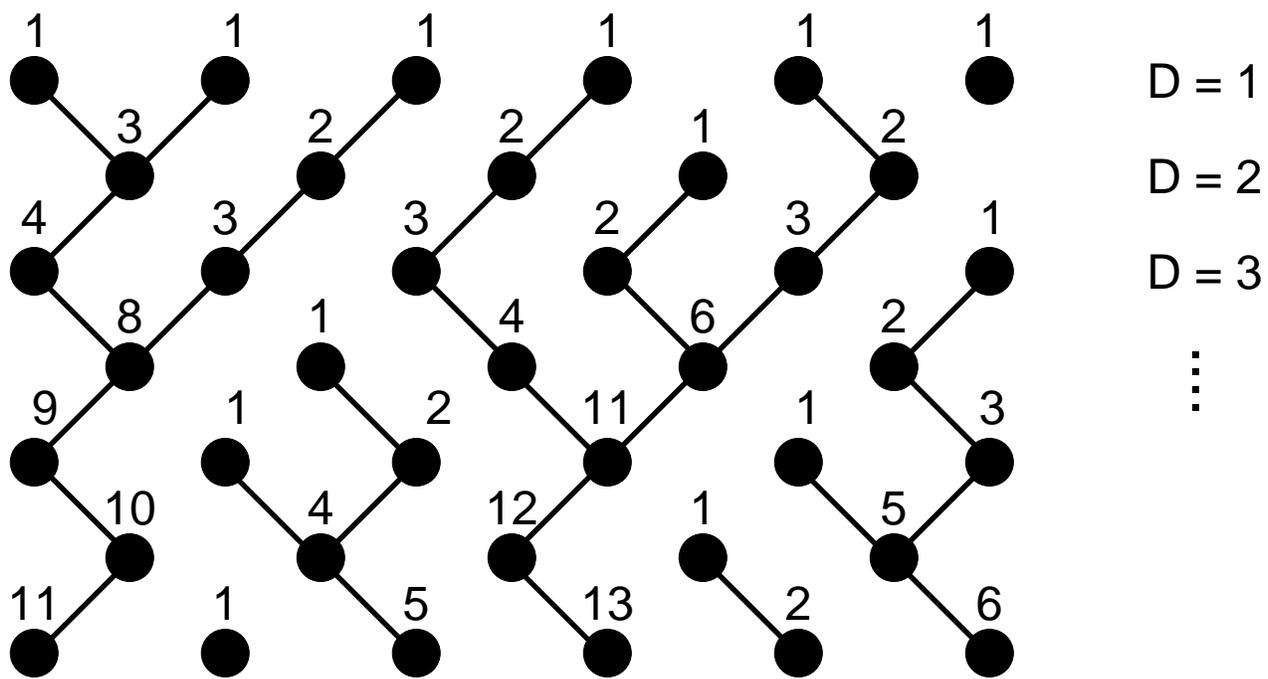
\begin{figure}
\vspace {1cm}
\caption{Schematic diagram showing the paths of weight
support for a two-dimensional system in the $q_{0,1}$
limit where each site transmits its weight to exactly
one neighbor below.
The numbers at each site are the values of $w(i,D)$.}
\label{fig:randomwalk}
\end{figure}

We first consider the case where each particle transmits
its weight to exactly one neighbor in the layer below,
so that the variable $q$ is restricted to taking on
only the values $0$ and $1$.
We denote this (singular) limiting case of our model by
the ``$q_{0,1}$ limit.''
Figure~\ref{fig:randomwalk} shows the paths of weight support
for a two-dimensional system in this limit.
The solid lines correspond to bonds for which $q=1$.
The paths of weight support of particles in the top row
are coalescing random walks.
Since a random walk of length $D$ has typical transverse
excursion of $D^{1/2}$, for the two-dimensional case the
maximum weight supported by an individual grain at depth
$D$ scales as $D^{3/2}$.\cite{Tak86}
Because $D^{3/2} \gg D$, the mean weight supported at depth
$D$, it is plausible
that in the $q_{0,1}$ limit the model yields a
broad weight distribution.

The defining equations of the $q_{0,1}$ limit of our model
are known to be identical to those of Scheidegger's model
of river networks\cite{Sch67}
and a model of aggregation with injection;\cite{Tak88,Hub91}
the model is also equivalent to that of the
directed Abelian sandpiles.\cite{Dha89,Maj93,Dhaunp}
(The number of neighbors below a particle, $N$, corresponds
to the dimensionality $d$ in these models.)
The last equivalence follows\cite{Dhaunp,Maj93} if we define
$G_0 ( \vec{X_1}; \vec{X_0} )$ as the probability that
the weight of site $\vec{X_1}$ is supported by site $\vec{X_0}$
in the same row or below it.
The conditional
probability that $\vec{X_1}$ is supported by $\vec{X_0}$,
given that $l$ of the $N$ neighboring particles
in the row below are supported by $\vec{X_0}$, is $l/N$.
Thus,
\be
G_0 (\vec{X_1};\vec{X_0}) = \frac{1}{N}\sum_{i=1}^{N}
G_0(\vec{X_1}-\vec{e_i};\vec{X_0}) + \delta_{\vec{X_1},\vec{X_0}}\ ,
\label{geqn1}
\ee
where $\{ \vec{X_1}-\vec{e_i}\}$ are the neighbors of $\vec{X_1}$
in the row below it, and the $\delta$-function term follows because
each particle must support its own weight.
Similarly, the probability that two sites $\vec{X_1}$
and $\vec{X_2}$ in the same row are supported by $\vec{X_0}$
satisfies:
\be
G(\vec{X_1},\vec{X_2};\vec{X_0}) = \frac{1}{N^2}
\sum_i \sum_j G(\vec{X_1}-\vec{e_i},\vec{X_2}-\vec{e_j};
\vec{X_0})
\label{geqn2}
\ee
for $\vec{X_1} \ne \vec{X_2}$.
These equations are precisely those that describe the
behavior of the correlations of the avalanches in the
directed Abelian sandpile.\cite{Dha89}\cite{sandpile}
In this model, an integer ``height'' variable $z(\vec{X})$
is assigned each site $\vec{X}$ on a lattice.
The dynamics are defined by the rule that if any
$z(\vec{X})$ exceeds a critical value, $z_c$, then
the variables at $m$ nearest neighbor sites along
a preferred direction increase
by 1, while $z(\vec{X})$ decreases by $m$.
In this context
$G_0(\vec{X_1}; \vec{X_0})$ is identified with the
probability that adding a particle
at $\vec{X_0}$ creates an avalanche that topples over
the site $\vec{X_1}$.
Higher order correlations are mapped similarly.
The distribution of weights in our model is mapped to the
distribution of avalanche sizes.

All these models\cite{Sch67,Tak88,Hub91,Maj93,Dha89}
have been studied as examples of self-organized
criticality,\cite{Bak87} because they lead to power
law correlations without an obvious tuning parameter.
However, in the context of our model, the $q_{0,1}$
limit is a singular one, where the probability of
$q \ne \{0,1\}$ has been tuned to zero.
As we shall show in this paper, generic distributions
$\eta(q)$, for which the probability that $q \ne \{0,1\}$
is nonzero (no matter how small), yield completely
different results, with the distribution of weights
decaying exponentially at large weights.
With hindsight, we identify the probability for a
river to split in the river network model,\cite{Sch67}
and the probability for a colloidal particle to
fragment in the aggregation model\cite{Tak88,Maj93}
as hidden parameters that were tuned to zero.
The corresponding parameter for directed Abelian
sandpiles is less obvious.

The equivalence of our model in the $q_{0,1}$ limit
to the models discussed above can be exploited to
obtain some results for the distribution of weights.
Recalling that the dimensionality, $d$, in these
models corresponds to our $N$, we know that the
weight distribution function at a depth $D$, $Q_D(w)$,
has a scaling form for all $N$:
\be
Q_D(w) = D^{-a}g( w/ D^{b} )\ ,
\label{wscaleeqn}
\ee
where $g(x) \rightarrow x^{-c}$ as $x \rightarrow 0$
(with a cutoff at $w$ of $O(1))$.

The normalization constraints,
$\int_0^{\infty} dw \ Q_D(w) = 1$ and
$\int_0^{\infty} dw\ w\ Q_D(w) = D$ yield the conditions
\be
a = b c,\ \ 1 + a = 2 b\ ,
\label{eq:exprelation1}
\ee
so that there is only one free exponent.
For $d=2$, the random walk argument at the beginning
of this subsection suggests that $b = 3/2$,\cite{Tak86}
which agrees with the exact result.\cite{Hub91}
For $d > 2$, random walks are less likely to coalesce,
and this argument breaks down.
In mean field theory one obtains the analytic result
$b = 2$,\cite{Tak88} and exact analytic results for directed
Abelian sandpiles in all dimensions\cite{Dha89}
show that mean field theory is valid for $d \ge 3$, and
confirm the result $b = 3/2$ for $d=2$.  (Our exponent $b$ can be
identified with $\alpha + 1$ of Ref.~\cite{Dha89}.)

As $D \rightarrow \infty$, the argument
of the scaling function $g$ in Eq.~(\ref{wscaleeqn})
is small for any finite $w$.
Thus, in the $q_{0,1}$ limit of our model,
the distribution of weights, $Q(w)$,
is independent of $D$ as $D \rightarrow \infty$,
and is of a power law form, and hence is infinitely
broad.


\subsection{The \q-model away from criticality.}
The rest of this paper concerns probability distributions
of the $q$'s that do not have the property that $q$ takes
on only the values $1$ and $0$.
We argue that all such distributions lead to
force distributions that differ qualitatively from
those described in the previous subsection.
The $q_{0,1}$ limit is the only
one that yields a power law force distribution;
other distributions lead to a much faster, typically
exponential, decay.
In addition, for other $q$-distributions, the
distribution for the {\em normalized} weight $v = w/D$,
$P_D(v)$ converges to a fixed distribution $P(v)$ as
$D \rightarrow \infty$.
In contrast, in the $q_{0,1}$ limit,
the quantity $Q_D(w)$ converges to a fixed function.
In this subsection we present evidence for these assertions
via both numerical simulations and mean field analysis.

\subsubsection{Numerical simulations}
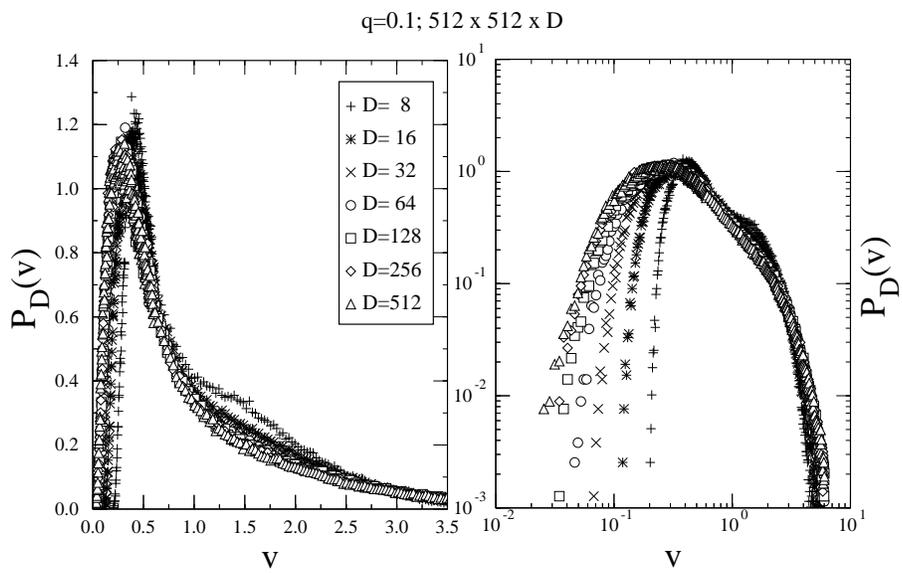
\begin{figure}
\vspace {1cm}
\caption{Linear-linear and log-log plots of the
normalized weight distribution function $P_D(v)$ versus
$v$ for a three-dimensional system on an fcc lattice
($N=3$), for the $q$-distribution defined in
Eq.~(\protect{\ref{eq:q0distdef}}) with $q_0 = 0.1$.
The distribution $P_D(v)$ appears to become independent
of $D$ as $D$ becomes large, and decays faster than
a power law at large $v$.}
\label{fig:q.1.eps}
\end{figure}

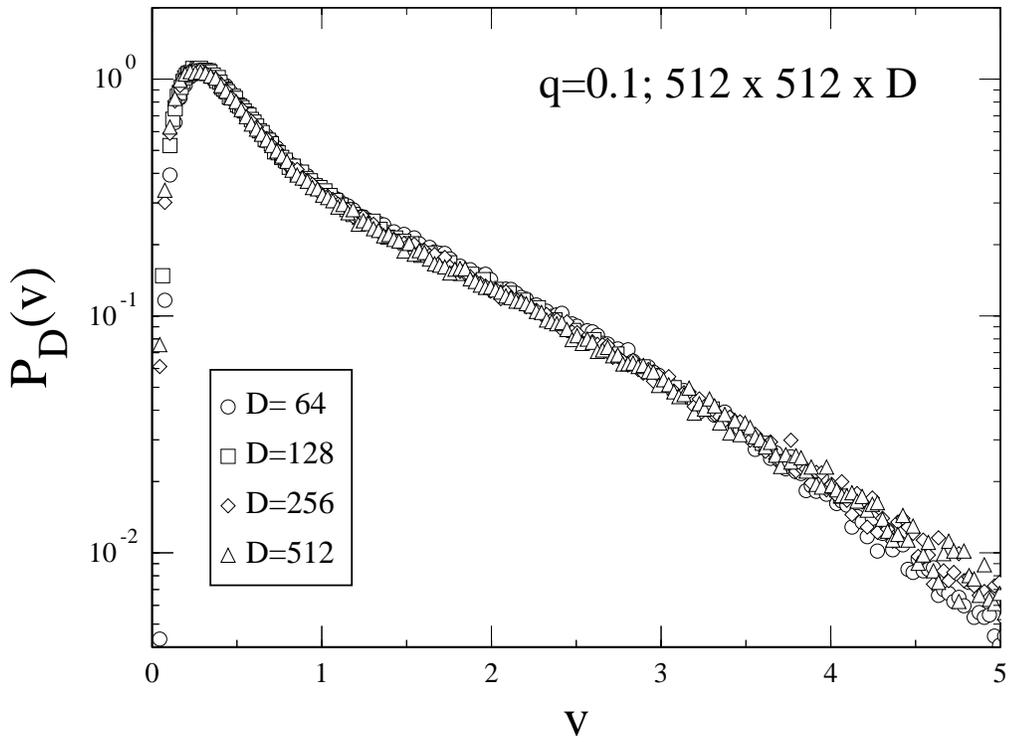
\begin{figure}
\vspace {1cm}
\caption{Semilog plot of the normalized weight distribution
function $P_D(v)$ versus $v$ for a three-dimensional system
on an fcc lattice ($N=3$), for the $q$-distribution defined
in Eq.~(\protect{\ref{eq:q0distdef}}) with $q_0 = 0.1$.
The behavior of $P_D(v)$ at large $v$ is consistent with
exponential decay.
}
\label{fig:q.1.semilog}.
\end{figure}

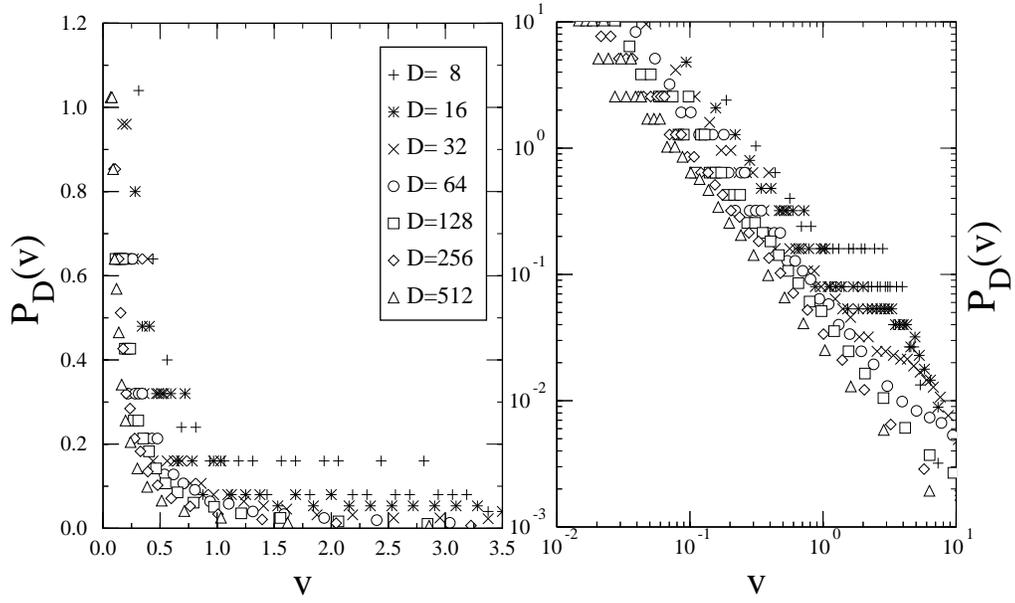
\begin{figure}
\vspace {1cm}
\caption{%
Linear-linear and log-log plots of the
normalized weight distribution function $P_D(v)$ versus
$v$ for a three-dimensional system on an fcc lattice
($N=3$), for the $q$-distribution defined in
Eq.~(\protect{\ref{eq:q0distdef}}) with $q_0 = 0$.
For this special case, the distribution $P_D(v)$ does not
become independent of $D$ as $D$ becomes large.
The asymptotic decay of $P_D(v)$ at large $v$ is a power law.
}
\label{fig:q0.eps}.
\end{figure}

Our numerical investigations
all indicate that that for all $q$-distributions
except for the $q_{0,1}$ limit, the normalized force
distribution $P_D(v)$ becomes independent of $D$
as $D \rightarrow \infty$.
To illustrate typical behavior, we consider the specific
$q$-distribution consisting of $N-1$ bonds emanating
down from each site with value $q = q_0 < 1/(N-1)$ and one bond
with $q = 1-(N-1)q_0$, which has the induced distribution:
\be
\eta_{q_0}(q) = \frac{1}{N} \delta(q-(1-(N-1)q_0))
+ \frac{N-1}{N} \delta(q-q_0)\ .
\label{eq:q0distdef}
\ee
Figure~\ref{fig:q.1.eps} displays the normalized force
distribution $P_D(v)$ versus $v$ for several different
depths $D$ in a 3-dimensional
fcc system ($N=3$) of dimension $512 \times 512 \times D$,
with $q_0 = 0.1$.
Periodic boundary conditions are imposed in the
transverse directions.
As $D$ becomes large, $P_D(v)$ converges to a function
independent of $D$ which decays faster than a power law.
Figure~\ref{fig:q.1.semilog} is a semilog plot of $P_D(v)$
versus $v$ for several values of $D$, showing that the
decay of $P(v)$ at large $D$ is roughly exponential.
To see that this behavior is qualitatively different from
that of the $q_{0,1}$ limit, in
Figure~\ref{fig:q0.eps} we display numerical results for
$P_D(v)$ versus $v$ for a system which is identical except that
$q_0 = 0$.
In contrast to the $q=0.1$ case, $P_D(v)$ decays as a
power law at large $v$.  Also, $P_D(v)$ shows no signs of
becoming independent of $D$ as $D \rightarrow \infty$.
This is consistent with the result in the previous section
that $Q_D(w)$ becomes independent of $D$ at large $D$.

\subsubsection{Mean field theory}
\label{sec:mft}

The technique of the mean field analysis for a general
$q$-distribution is a generalization of that used
for the $q_{0,1}$ case.\cite{Tak88}

The weight supported by a given site at depth $D$,
$w_i(D)$, depends not only
on the weight supported by the sites at depth $D-1$
but on the values of $q$ for the relevant bonds:
\be
w_i(D+1) = \sum_j q_{ij}w_j(D) + 1\ .
\label{modeldef}
\ee
In general the values of $w$ at neighboring sites
in layer $D$
are not independent; the mean field approximation consists
of ignoring these correlations.

As discussed above, when $q$ is allowed to take on values
other than $0$ and $1$, it is useful to study the force
distribution function as a function of the
{\it normalized} weight at a given depth, $v=w/D$.
In terms of the normalized weight variable $v$, the mean field
approximation leads to a recursive equation for
the weight distribution function $P_D(v)$:
\be
P_D(v) = \prod_{j=1}^{N}\{ \int_0^{1}dq_{j}
 \eta(q_{j})
 \int_0^{\infty}dv_j P_{D-1}(v_j) \}
 \delta(\sum_{j=1}^{N} [(D-1)/D]v_j q_j-(v-1/D))\ .
\label{Peqn}
\ee
The quantity $\eta(q) $ is defined in Eq.~(\ref{eq:etadef}).
The constraint that the $q$'s emanating downward from a
site must sum to unity enters only through the definition
of $\eta(q)$ because there
is no restriction on the $q$'s for the {\it ancestors} of a site.
The only approximation here is the neglect of possible
correlations between the values of $v$ among the ancestors.

By Laplace transforming, one finds that $\tilde{P}_D(s)$,
the Laplace transform of the distribution function
of the normalized weight $P_D(v)$, obeys:
\be
\tilde{P}_D(s) = e^{-s/D} \left[
 \int_{0}^1 dq \eta(q)
 \tilde{P}_{D-1}(s q (D-1)/D)\right]^N \ .
 \label{Ptildeeqn}
\ee
Since as $D \rightarrow \infty$ the distribution
$P_D(v)$ becomes independent of $D$,\cite{normnote}\cite{newnote}
Eq.~(\ref{Ptildeeqn}) then becomes:
\be
\tilde{P}(s) = \left[ \int_{0}^1 dq \eta(q)
 \tilde{P}(s q)\right]^N \ .
 \label{Preceqn}
\ee

First we show that the weight distribution $P(v)$ decays
faster than any power of $v$ for all $q$-distributions
except those that only take on the values $0,1$.
We expand the Laplace transform
$\tilde{P}(s)$ in powers of $s$,
$\tilde{P}(s) = 1+\sum_{j=1}^\infty P_js^j$,
and plug into Eq.~(\ref{Preceqn}), obtaining:
\be
1+P_1 s
+\sum_{j=2}^\infty P_j s^j\ =
[1+P_1 s/N +\sum_{j=2}^\infty P_j s^j \langle q^j\rangle]^N\ .
\ee
Here, $s$ is the Laplace transform variable,
$\langle q^j\rangle = \int_0^1 dq\ q^j \eta(q)$,
and we have used $\langle q \rangle = \frac{1}{N}$.
Equating the coefficients of $s^j$ on the left and
right hand sides of the equation, we obtain a linear
equation for $P_j$:
\be
P_j[N\langle q^j\rangle -1]=G(P_{j-1},P_{j-2},\ldots,P_{1})\ ,
\label{momenteqn}
\ee
where $G$ is some complicated polynomial.
This can be iterated to obtain $P_j$ for successively higher
values of $j$.

Since Eq.~(\ref{momenteqn}) is linear in $P_j$, $P_j$ can
diverge only if its coefficient, $[N\langle q^j\rangle-1]$, is zero.
If $q$ can take on only the values $0$ and $1$, then
$\langle q^j \rangle = \langle q \rangle$
and $[N\langle q^j\rangle-1] = 0$ for all $j>1$.
However, for ${\em any}$ other distribution of $q$'s
restricted to the interval $[0,1]$,
the distribution for $q^j$ is shifted towards
the origin compared to the distribution for $q^k$, whenever
$j>k$.
Since $\langle q^j\rangle < \langle q\rangle=1/N$ for all $j>2$,
Eq.~(\ref{momenteqn}) has a nonzero coefficient for
$P_j$ for all $j>2$, which means that all moments
$\langle v^j \rangle$ of $P(v)$ are finite.
(For the special case of $j=1$, the equation is
degenerate; $P_1$ is set by the normalization of $v$.)
If $\ln P(v)$ were to behave asymptotically as $-a\ln v$,
then $\langle v^j\rangle$ would diverge for all $j>a-1$.
Hence $P(v)$ must fall off faster than any power of $v$
and $(d \ln P(v)/ d \ln v) \rightarrow -\infty$
as $v \rightarrow \infty$.


\subsection{Weight distributions away from criticality:
Mean field results}

Now we consider the distribution of weights for non-critical
distributions of $q$'s.
Motivated by the geometrical disorder present in granular
materials, we focus especially on continuous distributions.
First we calculate this distribution within a mean field
approximation for the simplest possible continuous distribution,
$f(q_{ij}) = constant$, or
$\rho(q_{i1},\ldots,q_{iN}) = (N-1)! \delta(\sum_j q_{ij} - 1)$
(the uniform $q$-distribution).
We show that within mean field theory, all ``typical'' continuous
$q$-distributions lead to a
force distribution that decays exponentially at large weights.
We will show later that the mean field solution is {\it exact}
for a countable set of $q$-distributions, including the
uniform $q$-distribution.


\subsubsection{Mean field theory for the uniform distribution}
\label{sec:mftuniform}

One example of a $q$-distribution that can lead
to an exponentially decaying distribution of weights is
the ``uniform'' distribution of $q$'s,
for which the probability of obtaining the values
$q_{i1},\ldots,q_{iN}$ is
$\rho(q_{i1},\ldots,q_{iN}) = (N-1)!\delta(\sum_j q_{ij} - 1)$.
We show in Appendix A that this distribution induces
$\eta_u(q) = (N-1)(1-q)^{N-2}$.
Thus, for this $q$-distribution
in the limit $D \rightarrow \infty$ the mean field force
distribution is the solution to the self-consistent equation:
\be
\tilde{P}(s) =
 \left [\int_0^1 dq (N-1) (1-q)^{N-2}
 \tilde{P} (s q ) \right]^N\ .\label{qeqn}
\ee

First consider $N=2$.
For this case $\eta(q) = 1$, so Eq.~(\ref{qeqn})
becomes:
\be
\tilde{P} (s) = \left [\int_0^1 dq
 \tilde{P} (s q ) \right]^2\ .
\label{n2qeqn}
\ee
Letting $\tilde{V}(s) = (\tilde{P}(s))^{1/2}$
and $u = q s$, one obtains:
\be
s \tilde{V}(s) = \int_0^s d u
\tilde{V}^2(u)\ .
\ee
Differentiating with respect to $s$ yields:
\be
\tilde{V}(s) + s \frac{d\tilde{V}}{ds} =
\tilde{V}^2(s)\ ,
\ee
which can be integrated to yield
\be
\tilde{V}(s) = \frac{1}{1-Cs}\ .
\ee
The constant of integration $C$ is determined by
the definition of the mean,
$\int_0^\infty dv ~ v~P(v) = -\frac{d\tilde{P}}{ds}|_{s=0} = 1$.
Thus, $C = \frac{d\tilde{V}}{ds}|_{s=0} = -1/2$.
Hence one finds $\tilde{P}(s) = 4/(s+2)^2$
and $P(v) = 4 v e^{-2v}$.

This method can be generalized for all $N$.
Defining
$\tilde{V}_N(s) = (\tilde{P}_N(s))^{1/N}$,
inserting in Eq.~(\ref{qeqn}), and differentiating
$N-1$ times, one finds that $\tilde{V}_N(s)$
obeys the differential equation:
\be
\frac{d^{N-1}}{d s^{N-1}}(s^{N-1}\tilde{V}_N(s))
= (N-1)!\tilde{V}_N^N(s)\ .
\ee
A solution to this equation is
$\tilde{V}_N(s) = \frac{C}{s+C}$, where
$C$ is any constant.
This can be shown by induction:
Assume that
\be
\frac{d^{N-2}}{d s^{N-2}}
\left (s^{N-2}\frac{C}{s+C}\right )
= (N-2)!\left (\frac{C}{s+C}\right )^{N-1}\ .
\ee
Then
\begin{eqnarray}
\frac{d^{N-1}}{d s^{N-1}}\left (s^{N-1}\frac{C}{s+C}\right )
& = &
\frac{d^{N-1}}{d s^{N-1}}
\left ((s+C-C)s^{N-2}\frac{C}{s+C}\right )
 \\
& = & -C \frac{d}{ds} \left [
(N-2)! \left (\frac{C}{s+C}\right )^{N-1} \right ] \\
& = & (N-1)!\left (\frac{C}{s+C}\right )^{N}\ \ .
\end{eqnarray}
Since direct substitution can be used to show that
the identity  holds for $N=2$, it holds for all $N$.

The condition
$\frac{d\tilde{V}}{ds} = - \frac{1}{N}$
is satisfied when $C=N$.
Hence one finds the weight distribution:
\be
P_N(v) = \frac{N^N}{(N-1)!}v^{N-1}e^{-Nv}\ .
\label{Psoln}
\ee

The question of uniqueness of this solution is discussed below.


\subsubsection{Mean field asymptotic force distribution for generic
continuous $q$-distributions}

We now show that, within mean field theory, generic continuous
$q$-distributions lead to weight distribution functions $P(v)$
for the normalized weight $v$ which have the asymptotic
forms  $P(v) \propto v^{N-1}e^{-Nv}$ as $v \rightarrow \infty$
and $P(v) \propto v^{N-1}$ as $v \rightarrow 0$.

We consider $q$-distributions of the form
$\rho(q_{i1},\ldots,q_{iN})=\{ \prod_j f(q_{ij})\}
\delta(\sum_j q_{ij} - 1)$
(the uniform distribution is $f(q_{ij}) = constant$).
If $f(q_{ij})$ has a nonzero limit as $q_{ij} \rightarrow 0$,
and does not have a $\delta$-function contribution at
$q_{ij}=0$, then
phase space restrictions imply that the induced distribution
$\eta(q) \sim (1-q)^{N-2}$ for $q \rightarrow 1$.
This is because if a site receives
a fraction q of the weight from one of its predecessors, then
the fractions received by all the {\it other\/} successors of
that predecessor, $\{q_2\ldots q_N\}$ must add up to $1-q$. For
$q$ close to 1, this gives a phase-space volume of the order of
$(1-q)^{N-2}$.

To determine the large $v$ asymptotics of $P(v)$, we use
the result of Sec.~(\ref{sec:mft}) that $P(v)$ must fall
off faster than any power of $v$.
We write the $D \rightarrow \infty$ limit of
Eq.~(\ref{Peqn}) as:
\be
P(v) = \left \{ \prod_{j=1}^N \int_0^\infty dv_j F(v_j) \right \}
\delta (v-\sum_j v_j) \ , \label{PFeqn1}
\ee
\be
F(v_j) = \int_0^1 dq_j P(v_j/q_j) \eta(q_j)/q_j \ .
\label{PFeqn2}
\ee
Since $P(v)$ decays quickly (in particular, faster than $1/v$),
the apparent singularity near $q=0$ in Eq.~(\ref{PFeqn2})
is not really there.  The integral is dominated by $q \approx 1$.
This follows because
\begin{eqnarray}
P(v/q) &=& P(v) \exp \left[ \left .
-\frac{\partial \ln P(v/q)}{\partial q}
\right |_{q=1} (1-q) + \ldots \right]
\nonumber
\\
&=&P(v) \exp \left[ \left . \frac{v}{q^2}
\frac{\partial \ln P(v/q)}{\partial (v/q)}
\right |_{q=1} (1-q) + \ldots \right]
\nonumber
\\
&=&P(v) \exp \left[ \left .
\frac{\partial \ln P(u)}{\partial \ln u}
\right |_{u=v} (1-q) + \ldots \right] \ .
\end{eqnarray}
Since $\partial {\ln P(u)}/\partial \ln u \rightarrow -\infty$
as $u \rightarrow \infty$,
this expression becomes very small as $1-q$ increases.
Thus, for large $v$, since $\eta(q) \sim (1-q)^{N-2}$
for $q \approx 1$,
\be
F(v) \sim P(v)/ \left [ \left |
\frac{\partial \ln P(v)}{\partial \ln v} \right | \right ]^{N-1}\ .
\ee

Already it is clear that $P(v)$ for any generic $q$-distribution
has the same large-$v$ asymptotics as the uniform distribution,
since the asymptotics are determined entirely by the phase space
restrictions on $\eta(q)$ for $q \approx 1$.
This decay also can be demonstrated explicitly by assuming
faster and slower decays and showing inconsistency with
Eq.~(\ref{PFeqn1}).
If $P(v)$ were to decay faster than exponentially, then
the convolution in Eq.~(\ref{PFeqn1}) would be dominated by
the region where all the $v_j'$s are roughly equal.
But since $[P(v/N)]^N \gg P(v)$, Eq.~(\ref{PFeqn1})
cannot be satisfied.
On the other hand, if $P(v)$ were to decay slower than
exponentially, then the convolution would be dominated
by the region where one of the
$v_i$'s is $\approx v$ and the others are $O(1)$.
Eq.~(\ref{PFeqn1}) would then imply:
\be
P(v) \sim P(v)/ \left [ \left |
\frac{\partial \ln P(v)}{\partial \ln v} \right | \right ]^{N-1}\ .
\ee
Since the expression in square brackets diverges with $v$,
this is not possible either.
Thus one must have $P(v) = h(v) \exp[ - av]$, where $h(v)$
varies more slowly than an exponential.
Eq.~(\ref{PFeqn1}) then implies:
\be
h(v) = \left \{ \prod_{j=1}^N \int_0^\infty dv_j h(v_j)
/v^{N-1} \right \}
\delta (v-\sum_j v_j) \ . \nonumber
\ee
This is satisfied by $h(v) \sim v^{N-1}$, so that
\be
P(v) \sim v^{N-1} \exp [-av]
\label{mf_asymp_eqn}
\ee
for $v \rightarrow \infty$.

Hence we have shown that for generic continuous $q$-distributions,
within mean field theory
$P(v) \rightarrow v^{N-1}\exp(-av)$ as $v \rightarrow \infty$.

\subsubsection{Mean field theory for singular $q$-distributions.}

We have shown that all $q$-distributions which satisfy the condition
$\int_q^1 dq\ \eta(q) \sim (1-q)^{N-1}$ as $q \rightarrow 1$ have
a weight distribution within mean field theory that is of the
form $P(v) \sim v^{N-1} \exp [-av]$ for large $v$.
This condition on $\eta(q)$ is satisfied under fairly general
assumptions: one requires (1)
that the probability density for any $q_{ij}$ in
Eq.~(\ref{modeldef}) have a nonzero $q_{ij} \rightarrow 0$
limit and (2) that it not have a $\delta$-function contribution
at $q_{ij} = 0$.
However, as we shall see below, to compare the results of the
$q$-model to molecular dynamics simulations and to experiments
on real bead packs, it is useful to consider the case where there
is a finite probability for some of the $q_{ij}$'s to be zero,
which implies that the induced distribution $\eta(q)$ has a
$\delta$-function at $q=0$ (and in some cases at $q=1$).\cite{prodnote}
Such a choice for $\eta(q)$ is also useful in examining
the crossover from the critical $q_{0,1}$ limit to the smooth
$q$-distributions considered in the previous subsection.
We will see that $q$-distributions of this type lead to force
distributions $P(v)$ that decay exponentially, though with
different power laws multiplying the exponential than
for continuous $q$-distributions.

We first note that, when $\eta(q)$ has a finite weight at
$q=1$, it is impossible for $\tilde{P}(s)$ to diverge at any $s$.
The solutions of the form $\tilde{P}(s) \sim 1/(q+s/s_0)^N$
obtained in section~(\ref{sec:mftuniform}) were possible because,
in Eq.~(\ref{Preceqn}), the integral over $q$ reduces the
singularity, which is compensated by the exponentiation.
With a finite weight at $q=1$, close to a divergence
at $s_0$ one would have $\tilde{P}(s) \propto [\tilde{P}(s)]^N$,
which would be impossible as $s \rightarrow s_0$.

It is instructive to consider first a simplified version of
such singular $q$-distributions.
Let us consider the case of $N=2$, and assume that $\eta(q)$
has the form:
\be
\eta(q) = \frac{1}{2}(1-\theta)\{\delta(q)+\delta(1-q)\} +
\theta \delta(q-\frac{1}{2})\ ,
\label{a1}
\ee
with $0 < \theta < 1$.
This $\eta(q)$ satisfies the conditions $\int dq\ \eta(q) = 1$
and $\int dq\ q\ \eta(q) = 1/2$ for all $\theta$.
Eq.~(\ref{Preceqn}) then simplifies to:
\be
\tilde{P}(s) = \left[\frac{1}{2}(1-\theta) +
\frac{1}{2}(1-\theta)\tilde{P}(s) + \theta \tilde{P}(s/2)\right]^2 \ ,
\label{a2}
\ee
where we have used the fact that $\tilde{P}(0)=1$.
Eq.~(\ref{a2}) can be solved as follows:
for small $s$, we know that $\tilde{P}(s) = 1-s +O(s^2)$
(the coefficient of the linear term being fixed by the normalization
condition $\int dv\ v \ P(v) = 1)$.
Starting with a small negative value of $s$, where $\tilde{P}(s)$
is approximated as $1-s$, Eq.~(\ref{a2}) can be iterated to find
$\tilde{P}(2^ns)$ for $n = 1,\ 2, \ldots$ (the correct root of
the quadratic equation is chosen by requiring $\tilde{P}(s) = 1$
for $\tilde{P}(s/2) = 1$).
Eventually the result of this iteration scheme is complex rather
than real, signifying that $s$ is in a region where $\tilde{P}(s)$
has a branch cut.
It is easiest to find the origin $s_0$ of this branch cut by
adjusting $\tilde{P}(s_0/2)$ so that Eq.~(\ref{a2}) has a double
root for $\tilde{P}(s_0)$, and then iterating {\it backwards}
to obtain $\tilde{P}(s_0/2^n)$.
As $n \rightarrow \infty$, by matching on to the requirement that
$\tilde{P}(s) = 1-s$ for $s \rightarrow 0$, one can obtain $s_0$.
It is clear from Eq.~(\ref{a2}) that in the vicinity of $s_0$
$\tilde{P}(s)$ is of the form $\tilde{P}(s_0) + \alpha \sqrt{s-s_0}.$
This yields
\be
P(v) \sim v^{-3/2} \exp[-s_0v]\ \ \ \ {\rm for}\ v \rightarrow \infty\ .
\label{a3}
\ee
Although the power law prefactor is different from that in
Eq.~(\ref{mf_asymp_eqn}), there is still an exponential decay.

We now consider possible changes to Eq.~(\ref{a3}) from choosing
$\eta(q)$ of a more complicated form than Eq.~(\ref{a1}).
For any $\eta(q)$ of the form
\be
 \eta(q) = \sum_{i=0}^n c_i \delta(q-\lambda^i) +
  (1-\sum c_i)\delta(q)
\delta(q)
\ee
with $0 < \lambda < 1$, one can use the method outlined
above to find that $\tilde{P}(s)$ has a square-root branch cut
at some $s_0$.
This answer is not affected by making $n$ large, so long as
$c_0$ remains nonzero.
As $n \rightarrow \infty$, with all the $c_i$'s for $i > 0$
tending to zero, we can approach arbitrary continuous distributions
for $\eta(q)$ with $\delta$ functions at $q=0$ and $q=1$.

For $N>2$, Eq.~(\ref{a2}) is changed to a higher order equation.
This, however, does not generically change the results above.
Even for higher order equations, the degeneracy of the roots
generally occurs only pairwise, so that close to the point of
degeneracy the singularly ranging roots still have a square-root
singularity.
It will, however, be possible to find {\em non-generic} choices
for $\eta(q)$ that could result in an asymptotic form
$P(v) \sim v^{-(1 + \frac{1}{m})} \exp [-av]$ with $N \ge m \ge 2$.


\subsection{Beyond mean field theory}

\subsubsection{Proof that mean field theory is exact}
\label{sec:exactu}
In this section we prove that the mean field solution
presented in the previous subsection is an {\it exact}
solution of the model with the uniform
$q$-distribution for any $N$.

In general, the mean field theory presented above is not exact
because it does not account for the fact that
two neighboring sites in row $D+1$ both derive a fraction
of their weight from the same site in row $D$.
Suppose a site $j$ in row $D+1$ has $w(j)$ much larger than
the average value.
Then it is likely that the weight supported by an
ancestor $w(i)$ in row $D$ is larger than average also.
Because this ancestor transmits its weight to a neighboring
site in row $D+1$ as well, there is a ``correlation''
effect that creates a greater likelihood that in a given
layer sites supporting large weight are close together.
On the other hand, there is a ``anticorrelation'' effect
arising because $\sum_j q_{ij} = 1$;
if a large fraction of the weight from site $i$ is
transmitted to site $j$, then small fractions are
transmitted to the other ``offspring'' sites.
When the $q$'s are chosen from the uniform distribution,
these ``correlation'' and ``anticorrelation'' effects
cancel exactly.

The result that the mean field correlation functions are
exact for the uniform distribution of $q$'s can be understood
by considering the system in terms of weights on bonds.
Each bond $\{ij\}$ corresponds to a particle with ``energy''
$E_{ij}=v_iq_{ij}$.
Moving down by one layer corresponds to having groups of $N$
particles colliding at each site and emerging with different
energies, subject to the constraint that the total energy
of all $N$ particles colliding at each site is unchanged by
the collision.

For the ``uniform'' $q$-distribution,
each collision takes $N$ particles of energies
$e_{\alpha_1}, \ldots, e_{\alpha_N}$ and changes their
energies to $E_{\alpha_1}, \ldots, E_{\alpha_N}$, subject only
to the constraint that $\sum e_\alpha = \sum E_\alpha$.
If we start with a ``microcanonical'' ansatz for the phase-space
density, i.e. that it is uniform over the space $\sum E_\alpha = E$,
then it is preserved by the collisions.
Hence, the microcanonical density is the correct one
for this system.

With a microcanonical density for a large collection of
particles, the density for any finite subgroup is
canonical (in the thermodynamic limit).\cite{Katz67}
Thus, we have shown for this case that the distribution
of ``bond forces'' is exponential,
which is the most random distribution consistent with
the constraint that the sum of the forces is
fixed.\cite{Katz67,Roth80}

Note that this argument does not hold for $q$-distributions
other than the uniform one.
For instance, in the $q_{0,1}$ limit,
each collision takes all the energy of the group and gives
it to one of the colliding particles.
Thus, even if we start with the microcanonical distribution,
it breaks down at the very first step.
For general $q$-distributions, the phase space density is
not separable, i.e., mean field theory is not exact and
there are spatial correlations within each layer.

The explicit algebraic proof proceeds by constructing
exact recursion relations for the
correlation functions describing the weight distribution
in the model in row $D+1$ in terms those for row $D$,
and showing that the mean field
correlation functions are invariant under this recursion.
We ignore the weight added in each row because we are
looking for the fixed distribution very far down the pile.

Let $P_D(u_i)$ be the probability that site $i$ in row $D$
supports weight $u_i$, $P_D(u_{i_1}, u_{i_2})$
be the probability that sites $i_1$ and $i_2$ support
weight $u_{i_1}$ and $u_{i_2}$, respectively, and
$P_D(u_{i_1},u_{i_2}, \ldots, u_{i_n})$ be the normalized joint
distribution describing the probability that sites
$i_1, i_2, \ldots, i_n$ support weights $u_{i_1}, \ldots, u_{i_n}$,
respectively.
The mean field joint probability distributions are given by the
mean field $P(u)$ and
\be
P(u_1,u_2,\cdots,u_n) = P(u_1)P(u_2)\cdots P(u_n)\ .
\label{mfcondition}
\ee

Consider the $M$-point correlation function in row $D+1$
that is obtained when all the correlation functions in row $D$ are the
the mean field ones.
Let $\{u_i\}$ be the weights in row $D$ and
$\{v_i\}$ be the weights in row $D+1$.
Consider a cluster of sites $j=1, \ldots, M$ in row $D+1$,
with ancestors in row $D$ at sites $i=1, \ldots, p$.
(The labels do not imply any particular spatial relation of the
sites.)
The $q$'s describing the bonds emanating from ancestor ${i}$
are $q_{i l}$, where $l = 1, \ldots, N$.
We define $\eta_{il}(j)$ to be $1$ if sites $i$ and $j$
are connected by bond ${il}$ and zero otherwise.
The $M$-point correlation function in row $D+1$,
$P_{D+1}(v_1, \ldots, v_M)$, must obey:
\begin{eqnarray}
P_{D+1}(v_1, \ldots, v_M) & = &
\prod_{i=1}^p \Big\{ \int_0^1 dq_{i1} \ldots
\int_0^1 dq_{iN} ~(N-1)!~ \delta(1-\sum_{k=1}^N q_{ik})
\nonumber
\\*
&~&~~~~~~\int_0^\infty du_i P_D(u_i)\Big\}
\prod_{j=1}^M \delta \left (v_j - \sum_{i=1}^p\sum_{l=1}^N
\eta_{il}(j) q_{il}u_i \right ) \ .
\label{2.37}
\end{eqnarray}

We define the general Laplace transform
\be
\tilde{P}(s_1,\ldots,s_n) = \int_0^\infty dv_1\cdots
\int_0^\infty dv_n P(v_1,\ldots,v_n)
e^{-s_1 v_1 \cdots -s_n v_n} \ .
\ee
Laplace transforming Eq.~(\ref{2.37}), one obtains:
\begin{eqnarray}
\tilde{P}_{D+1}(s_1, \ldots, s_M) & = &
\prod_{i=1}^p \int_0^1 dq_{i1} \ldots
\int_0^1 dq_{iN} ~(N-1)!~ \delta(1-\sum_{k=1}^N q_{ik})
\nonumber
\\*
&~&~~~~~~~~~~
\tilde{P}_D\left (\sum_{j=1}^M\sum_{l=1}^N
\eta_{il}(j) q_{il}s_j\right ) \ .
\end{eqnarray}
For $\tilde{P}_D(x) = (1 + x/N)^{-N}$,
one can use the condition $\sum_{l=1}^N q_{il} = 1$
to write:
\begin{eqnarray}
\tilde{P}_{D+1}(s_1, \ldots, s_M) & = &
\prod_{i=1}^p \int_0^1 dq_{i1} \ldots
\int_0^1 dq_{iN} ~(N-1)!~ \delta(1-\sum_{k=1}^N q_{ik})
\nonumber
\\*
&~&~~~~~~~~~~
\left (\sum_{l=1}^N q_{il}
(1+\sum_{j=1}^M
\eta_{il}(j) s_j/N)\right )^{-N} \ .
\end{eqnarray}
Using the identity\cite{zinn90}
\be
\prod_{n=1}^N (a_n)^{-1} = (N-1)!
\int_0^1 dx_1 \ldots \int_0^1 dx_N \frac{\delta(1-x_1 - \ldots-x_N)}
{(a_1 x_1 + \ldots + a_N x_N)^N}
\label{identityn}
\ee
with $a_n  = 1+\sum_{j=1}^M \eta_{in}(j)s_j/N$,
one finds:
\be
\tilde{P}_{D+1}(s_1, \ldots, s_M) =
\prod_{i=1}^p\prod_{n=1}^N
\frac{1}{1+\sum_{j=1}^M\eta_{in}(j)s_j/N}\ .
\label{product}
\ee
If a given bond $\{in\}$ connects to no sites in the
descendant cluster, then every term in the sum in the
denominator of Eq.~(\ref{product}) is zero, and the
$\{in\}^{th}$ term in the product is unity.
If the bond connects to a site in the descendant cluster, then
$\eta_{in}(j)$ is unity for exactly one $j$.
Each site $j$ in the descendant cluster is connected to exactly
$N$ antecedents in row $D$, so:
\be
\tilde{P}_{D+1}(s_1, \ldots, s_M) = \prod_{j=1}^M
\frac{1}{(1+s_j/N)^N}\ .
\ee
Thus, the mean field correlation functions are preserved
from row to row for this $q$-distribution.

\subsubsection{Other $q$ distributions}
We have identified a countable set of $q$-distributions
for which mean field theory is exact, those of the
form $f(q_{ij}) = q^r$, for all integer $r$
(the uniform distribution is $r=0$).
The resulting force distribution
$P_r(v) \propto v^{r+N-1}e^{-Nrv}$ has Laplace
transform $\tilde{P}_r(s) = (1+s/(Nr))^{-Nr}$.
The demonstration that this solution is exact follows precisely
the same line of reasoning as for the $r=0$ case presented
in the previous subsection, utilizing the identity:\cite{zinn90}
\be
\prod_{n=1}^N (a_n)^{-r} = \frac{\Gamma(Nr)}{[\Gamma(r)]^N}
\int_0^1 dx_1 x_1^{r-1}\ldots \int_0^1 dx_N x_N^{r-1}
\frac{\delta(1-x_1 - \ldots-x_N)}
{(a_1 x_1 + \ldots + a_N x_N)^{Nr}} \ .
\ee

In terms of the particle collision picture discussed at the
beginning of Sec.~(\ref{sec:exactu}), a general value of $r$
corresponds to the particles having an energy which is the
sum of $r+1$ components (which may be viewed as as spatial
coordinates in some underlying space), each one of which is
conserved individually in a collision.
The microcanonical phase-space density, uniform in the
$N(r+1)$-dimensional space, is preserved by the collisions,
and yields the $P_r(v)$ state here.

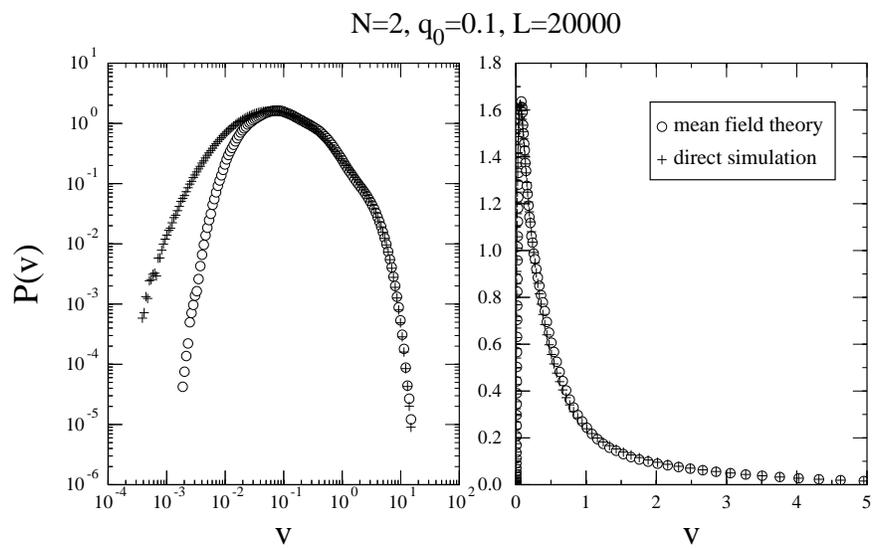
\begin{figure}
\vspace {1cm}
\caption{Comparison of force distribution $P(v)$ versus
$v$ for simulation of mean field theory and of the original
model equations (\protect{\ref{modeleqn}}) for a system
with the $q$-distribution Eq.~(\protect{\ref{eq:q0distdef}})
with $N=2$ and $q_0=0.1$.
Both data sets were obtained by averaging the bottom $10000$
rows of a $20000 \times 20000$ system.
Mean field theory does not yield the exact $P(v)$ for this
$q$-distribution.
Nonetheless, it provides an accurate quantitative estimate
for $P(v)$ over a broad range of $v$.
}
\label{fig:mftnotexact}
\end{figure}

The result that mean field theory yields an exact solution
of the model holds only for a very limited class
of $q$-distributions.
For general $q$-distributions, the phase space density is
not separable, i.e., mean field theory is not exact and
there are spatial correlations within each layer.
For example, Figure~\ref{fig:mftnotexact} shows $P(v)$ for
a two-dimensional system with $N=2$ and the $q$-distribution
where the two bonds emanating from each take on the values
$q_0$ and $1-q_0$, with $q_0 = 0.1$.
In the model, a site $(i,D)$ is connected to sites
$(i,D+1)$ and $(i+1 ~mod~L,D+1)$;
in the mean field calculation, site $(i,d)$ is connected
to sites $(p_1(i),D+1)$ and $(p_2(i),D+1)$, where
$\vec{p_1}$ and $\vec{p_2}$ are permutations of
$(1,\ldots,L)$.
This method of simulating mean field theory destroys the
spatial correlations between ancestor sites, while ensuring
that every site has exactly two ancestors and two descendants.
The numerical data were obtained by averaging
$P(v)$ for rows $10,001-20,000$ in a system of transverse
extent $L=20,000$.
This figure demonstrates explicitly that the mean field
force distribution $P(v)$ is not exact for this distribution.
However, the deviations of the mean field theory from the
direct simulation are extremely small for $v \simgt 0.1$,
so mean field theory provides an accurate quantitative
estimate for $P(v)$ over a large range of $v$.

\subsubsection{Uniqueness of the steady state distribution}

In this subsection we show that our results (numerical and analytical)
for the force distribution
do not depend on either the boundary conditions imposed
at the top of the system or on the specific realization
of randomness a particular system might have.

Consider a system of finite transverse extent $L$, in which weights
$\{w(i)\}$ are put on the particles in the top row.
The weight then propagates downwards according to Eq.~(\ref{modeldef}).
If we now consider the same system with a different loading on
the top row, $\{w(i) + \delta w (i)\}$, then
since Eq.~(\ref{modeldef}) is linear in $w$, the
difference between the two solutions satisfies the
homogeneous equation
\be
\delta w (D+1,j) = \sum_i q_{i,j}(D) \delta w(D,i)\ .
\label{b1}
\ee
Summing up both sides, we see that
$\sum_j \delta w (D+1, j) = \sum_i \delta w(D,i)$,
which means that the total excess weight placed on the
top of the system progagates downwards unaltered.
Such a change only affects the normalization of
our distributions.
Thus, if we are interested in the normalized distribution
$P_D(v)$, we can without loss of generality consider
perturbations $\{\delta w (D,i)\}$ satisfying the
constraint $\sum_i \delta w (D,i) = 0$.

Equation~(\ref{b1}) can be viewed as a two stage process:
(1) each $\delta w(D,i)$ splits into $N$ parts,
$q_{i,j}(D) \delta w(D,i)$, and (2) the $N$ fragments
$q_{i,j}(D) \delta w(D,i)$, with $i$ running over
the neighbors of $j$ in the row above it, combine to
give $\delta w(D+1,j)$.
The important thing is that all the $q_{i,j}$'s are positive.
Thus if we define the total difference between the two
configurations as $\Delta(D) = \sum_i |\delta w(D,i)| $,
then because all the $q$'s are positive and $\sum_j q_{ij} = 1$,
$\Delta$ is unchanged in the first step, while
in the second step it can either stay constant or decrease
(depending on whether the signs of the fragments are the
same or different).
Further, while for any {\it particular} value of $D$ it is
possible for $\Delta(D)$ to be equal to $\Delta(D+1)$,
the only way in which $\Delta(D)$ can remain unchanged
as $D$ increases is if all the positive $\delta w$'s are
segregated from all the negative ones.
Even if this is the case in the top row, this becomes
increasingly unlikely as $D$ is increased.
In fact, if the minimum distance between positive
and negative $\delta w$ is $l$ in the top row, and if
there are no $\delta$-functions in $\eta(q)$, then
$\Delta(D+1)$ {\it must} be less than $\Delta(D)$
for $D>l$.
Thus for a system of finite transverse extent, the
distribution of weights at the bottom of the system
is independent of the loading on the top row in the
limit that the height of the system is infinite.
For the case when all dimensions of the system are made
infinite the situation is trickier; due to the conservation
of $\sum \delta w$ under the evolution of Eq.~(\ref{b1})
discussed above, if one were to make $\delta w$ positive
on one side of the top row and negative on the other
half, for a system of transverse extent $L$ it would
require a height $O(L^2)$ for the
effects of this perturbation to ``diffuse'' away.
For generic loading at the top, however, we do not
expect such an anomalous concentration of fluctuations
into only the longest wavelength modes of the system,
and $\Delta(D)$ should decay with $D$ even if all
dimensions of the system are enlarged.

We have seen that the distribution of weight at
the bottom of any infinite system is independent of the
details of how forces are distributed at the top, at
least in the limit when the height of the system is
taken to infinity before its transverse dimensions.
This is true for each system individually, and is
therefore also true for the full ensemble of systems
with different realizations of randomness (the choice
of $q_{ij}$'s), so that the solutions we have obtained
so far for quantitites like $P(v)$ are unique.
For any particular system, however, the weights on the
different sites at the bottom {\it do} depend on
the $q_{ij}$'s; in fact, with all the $q_{ij}$'s
specified, the weights on the different sites are
completely determined.
Even for a single system, however, statistics can be
obtained by measuring quantities across all the sites
in the bottom row; for a system of infinite transverse
size the measurements then lead to distributions.
At least for the ``uniform'' $q$-distribution,
any quantity like, say, $P(v)$, is the same, whether
obtained by averaging over sites in a single system
or for a single site over the entire ensemble.
This is because, as we have seen, the ensemble
averaged distribution of $\{v_1, \ldots, v_L\}$ across
the bottom row is of the form
$P(v_1, \ldots, v_L) = P(v_1)P(v_2)\ldots P(v_L)$.
For any single system chosen randomly from the
ensemble, this is the probability density that
the normalized weights in the bottom row take on
the specific values $\{v_1,v_2, \ldots , v_L\}$.
The probability that $l$ of these $L$ sites
will have $v_i$ greater than some $v_0$ is then
\be
\left( \begin{array}{c}
L \\l
\end{array}
\right)
\left(\int_{v_0}^\infty P(v) dv \right) ^l
\left( 1 - \int_{v_0}^\infty P(v) dv \right)^{L-l}\ ;
\ee
as $L \rightarrow \infty$, $l/L$ is sharply peaked
around $\int_{v_0}^\infty P(v) dv$, so that the
site averaged result is the same as the ensemble average.
We expect this to be the case even for more general
$q$-distributions, for which the ensemble averaged
$P(v_1,v_2,\ldots,v_L)$ does not have a product form,
so long as the transverse correlation lengths are finite.


\section{Numerical simulations of sphere packings}
\label{sec:simulation}
We now discuss the relevance of the $q$-model to granular materials.
Although we have shown that the $q$-model yields an
exponentially-decaying force distribution independent of the
details of the $q$-distribution, to
make quantitative comparison of this model to granular systems,
we must know the $q$-distribution for a granular material.
To make this comparison,
we have performed molecular dynamics simulations of three-dimensional
sphere packings, analyzed the contact distributions to estimate the
distribution of $q$'s, and
then calculated the force distribution in the sphere packing
and compared it to that predicted by the $q$-model.
Our simulations yield results for the contact force distributions
that are consistent with previous work;\cite{Cundall89,Jenkins89,%
Bagi93,Pav94}
the new ingredient here is that the geometry of the packing
is characterized simultaneously, allowing testing of the
statistical assumptions underlying the $q$-model.

Our simulation consists of 500 spherical beads of weight and diameter
unity in a uniform gravitational field with gravitational constant
$g=1$, interacting via a
central force $F$ of the Hertzian form, $F = F_o (\delta r)^{3/2}$.
Here, $F_o$ is the force constant, chosen so that
a sphere has a deformation of $\delta r = 0.001$ when
subjected to its own weight, and $\delta r$ is the
deformation of each bead at the contact.
The box containing the beads had a fixed bottom,
and lateral dimensions of $5.5\times 5.5$.
In each simulation, the spheres are initially placed in a loose
rectangular lattice with lattice constants of $1\times 1\times 1.5$
and have random initial velocities uniformly distributed in the range
$-V_{max}<V_{x,y,z}<V_{max}$, where $V_{max}= 50$ is large enough to
yield significantly different packings from run to run.
By freezing the motion of the beads whenever the total kinetic energy of
the system reaches a maximum, the kinetic energy of the system is reduced
and eventually the spheres all settle to the bottom of the box.
Starting with a flat bottom, the regularity of layer-like packing reduces
as the height increases.  A rough bottom was obtained by selecting the
beads with height between $H$ and $H+1$ (typically, $H \sim 10$)
and this rough bottom was used for the next simulation.
Within a few iterations, the statistical properties of the rough bottom
becomes independent of its initial configuration; this configuration
of spheres at the bottom of the box is then fixed and used as a boundary
condition for subsequent packing simulations.

In our packings, a sphere can have up to 6 contacts
on its bottom half.
However, on average, the three strongest vertical forces at
these contacts sustain over 98\% of the load;
three or fewer particles supported at least
$90\%$ of the weight for over $92\%$ of the particles.
Therefore, comparison with the $q$-model
with $N=3$ is reasonable.

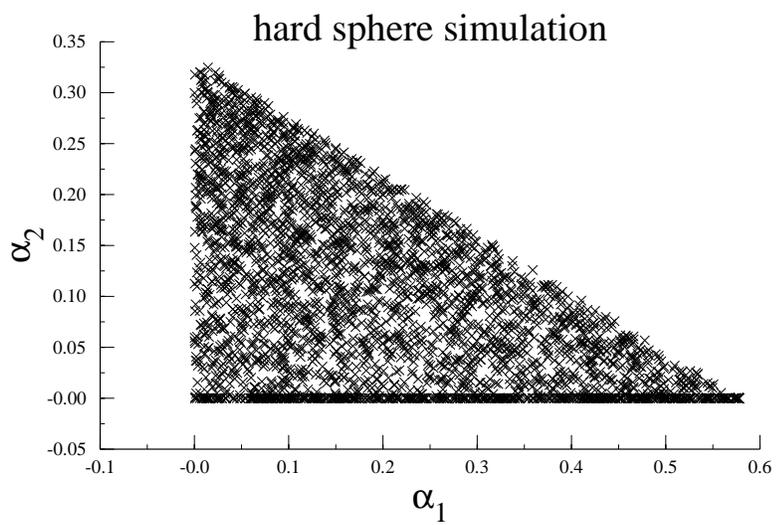
\begin{figure}
\vspace{1cm}
\caption{Scatter plot of contact variables $\alpha_1$,
$\alpha_2$ (defined in the text) obtained from the
sphere simulation described in the text.
The graph has 3229 points.
On this plot, the uniform $q$-distribution would have
a uniform density of points.
The ``arching'' in the simulation is reflected in the fact
that a nonzero fraction of points have $\alpha_2=0$.}
\label{fig:newtriangle},
\end{figure}

We estimate the $q$-distribution for the sphere simulation
by calculating the fractions of the total vertical force
supported by each of the three strongest
contacts.\cite{threenote}
To display our results for the $q$-distribution for
the simulation of hard spheres, we define the variables
$\alpha_1=(q_3-q_2)/\sqrt{3}$, $\alpha_2=q_1$.\cite{alphanote}
Because $\sum_{j=1}^3q_j = 1$, the possible values of
the $\alpha's$ can all be represented as points in the
interior of an equilateral triangle, where
the values of $q$ are the perpendicular distances to
each side of the triangle.
Moreover, for the uniform $q$-distribution, the
density of points in the triangle is constant.
If one orders the $q$'s so that $q_3>q_2>q_1$, then
in terms of the $\alpha$ variables, all the points
lie in the triangle shown in Figure~\ref{fig:newtriangle},
which is bounded by the lines $\alpha_1>0$, $\alpha_2 > 0$,
and $\sqrt{3}\alpha_1+\alpha_2<1$.
As Figure~\ref{fig:newtriangle} demonstrates, there is
some deviation from the uniform $q$-distribution because
a nonzero fraction of the particles have $q_1=\alpha_2=0$.
A reasonable description of the numerically observed
particle contact distribution is
obtained by taking each particle and assigning with
probability $p$, $l$, and $u$ into ``point'', ``line'',
and ``uniform'' pieces.
In the ``point'' piece one of the $q$'s has value unity,
and the other two are zero.
In the ``line'' piece one of the $q$'s is set to zero,
and the other two are determined as in the $N=2$
uniform distribution.
Finally, the particles in the ``uniform'' piece have
their $q$'s determined exactly as in the $N=3$
uniform distribution.
Our numerical data for the spheres are consistent with
the values $p=0.017 \pm 0.0023$, $l=0.1635 \pm 0.007$,
$u=1-l-p=0.8195 \pm 0.007$.

We now discuss our results for the distribution of vertical
forces.
First, we calculated the force distribution
at several different depths $D$.
Our numerical data indicate that if one considers
the normalized force $v=w/D$, the force distribution $P(v)$
indeed becomes independent of depth for $D \simgt 5$, and it
decays exponentially at large $v$.
The data were obtained by making a histogram of the vertical
force exerted by spheres in horizontal slices of width $\Delta D=1$.
The scales are set by the normalization requirements
$\int_0^\infty dv\ P(v) = 1$,
$\int_0^\infty dv\ v\ P(v) = 1$.

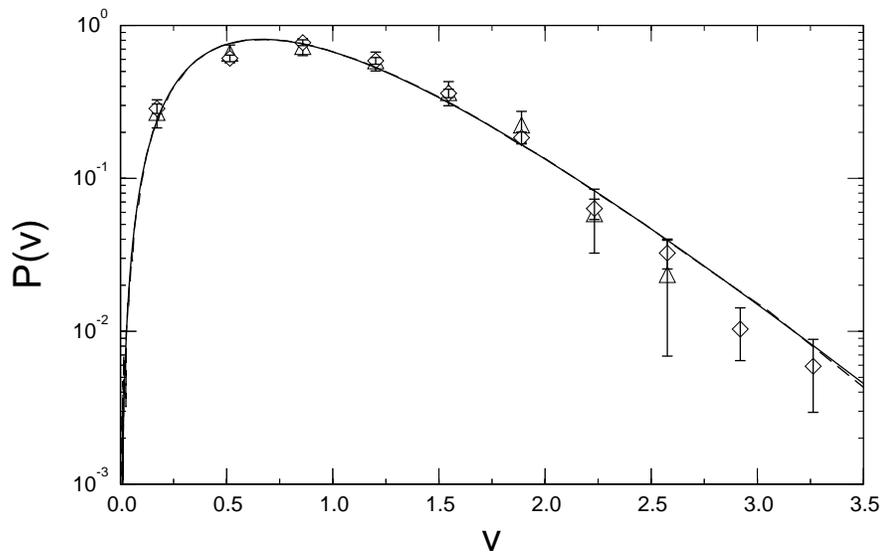
\begin{figure}
\vspace {1cm}
\caption{The distribution of forces $P(v)$ as a function of
normalized weight $v = w/D$ at a given depth $D$.
Dashed line:  $P(v)$ at $D=1024$ obtained via numerical simulation
of the model Eq.~(\protect{\ref{modeleqn}})
with $f(q) = 1$ on a periodically
continued fcc lattice of transverse extent 1024.
Solid line:  $P_u(v)$ obtained from the analytic mean field solution,
Eq.~(\protect{\ref{Psoln}}).
The points are $P(v)$ obtained in the sphere
simulation described in the text at depth $D=10$ (triangles)
and averaged over depths $D=6$ through $D=13$ (diamonds).
There are no adjustable parameters; the scales are set by
the normalization requirements $\int_0^\infty dv\ P(v) = 1$,
$\int_0^\infty dv\ v\ P(v) = 1$}
\label{fig:pofv1}
\end{figure}

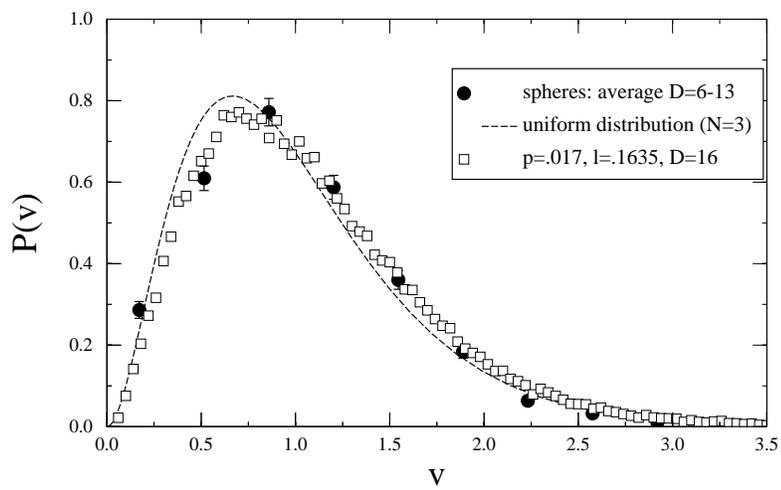
\begin{figure}
\vspace {1cm}
\caption{The distribution of forces $P(v)$ as a function of
normalized weight $v = w/D$ at a given depth $D$.
Dashed line:  $P_u(v)$ obtained from the analytic mean field solution
for $q$-model with $N=3$, Eq.~(\protect{\ref{Psoln}}).
Solid circles:  $P(v)$ obtained in the sphere simulation
averaged over depths $D=6$ through $D=13$.
Open squares:  $P(v)$ at $D=16$ obtained via numerical simulation
of the model Eq.~(\protect{\ref{modeleqn}})
with the three-part $q$-distribution described in the
text, with parameter values given on the graph, on a periodically
continued fcc lattice of transverse extent 256.
This figure demonstrates that using the measured $q$-distribution
instead of the uniform $q$-distribution
improves the already good agreement between the $q$-model and
the sphere simulation.}
\label{fig:newcomp}
\end{figure}

We now compare the results of these molecular dynamics
simulations to results from the $q$-model.
Figure~\ref{fig:pofv1} shows $P(v)$ calculated via numerical
simulation of the $q$-model Eq.~(\ref{modeleqn}) with $f(q) = 1$
at depth $D=1024$ on a periodically continued fcc latice of
side $1024$, with $N=3$.
Within our approximation of placing the grains on a
uniform lattice, the reasonable choice for $N$ is
the dimensionality $d$ of the system:  For $d=3$ the
grains are approximated as being in triangular lattice
layers, with each layer staggered relative to the
next, so that each grain has three neighbors.
As expected, since the mean-field distribution is exact for
the uniform case, there is excellent agreement with Eq.~(\ref{Psoln}).
On the same graph we show $P(v)$ obtained in the sphere
simulation described above.
Both the sphere simulation and the $q$-model exhibit
a $P(v)$ that decays exponentially at large $v$.
The quantitative agreement between the two is surprisingly good
considering the ``arching''\cite{sandrefs} in the sphere simulation,
as reflected in the ``line'' and ``point'' pieces of the
$q$-distribution for the spheres.
To examine the effects of arching on the results, we examined
the force distribution resulting from the ``$q$-model'' with
the three-piece $q$-distribution, which more closely
approximates that of the sphere simulation.
Figure~\ref{fig:newcomp} shows the numerically calculated $P(v)$
for the $q$-model with the three-piece distribution with $p=0.017$
and $l=0.1635$, together with the solution for the uniform
distribution and the numerical data from the sphere simulation.
Changing the $q$-distribution has little effect on $P(v)$;
to the extent that there is a change, it appears to improve
the already good agreement between the $q$-model and the
sphere simulation.

Thus, our simulations indicate that our sphere packings are
reasonably well-described (at the $\sim 15\%$ level) by the
uniform $q$-distribution.
Deviations from this $q$-distribution are observed;
accounting for them improves the already good agreement
between the $q$-model and the simulations.

\section{Discussion}
This paper presents a statistical model
for the force inhomogeneities in static bead packs
and compares the results to numerical simulations of
disordered sphere packings.
The irregularities of the packing are described probabilistically,
in terms of spatially uncorrelated random variables.
Although there is a special $q$-distribution for the $q$-model
that leads to a force distribution that
decays as a power law at large forces,
we have presented evidence that the force distribution
decays exponentially at large forces for almost all
$q$-distributions.
We obtain exact results for all the multipoint force
correlation functions at a given depth for a countable
set of $q$-distributions, including one that is ``generic''
(the ``uniform'' distribution).
The force distribution function for the uniform case agrees
quantitatively with that obtained for the sphere simulation.
Our numerical calculations demonstrate that a modified
distribution of $q$'s which more closely approximates that
observed for the sphere simulation improves the already
good agreement between the force distribution predicted
by the $q$-model using the uniform $q$-distribution and
simulations of spheres.
Thus, this model appears to contain some essential features
of the force inhomogeneities in granular solids.

Neither our simulations nor the q-model of Eq.~(\ref{modeleqn})
captures all features of real bead packs.
In our simulations, we have included only central forces and
have ignored friction;
the q-model ignores the vector nature of the forces, assuming
that only the component along the direction of gravity
plays a vital role.
The qualitative consistency between the results obtained
using the different methods as well as with experiment\cite{liu95}
provides some indication that the
effects that we have neglected do not determine the main
qualitative features of the force distribution at large $v$.

Several avenues for future investigations are evident.
It should be straightforward to extend the analysis of the
model to calculate longitudinal (along the direction of
gravity) correlations of the forces.
It is not obvious how to measure these correlations experimentally,
but comparison to sphere simulations is clearly possible and
would provide further tests of the statistical model.
Similarly, the theory makes clearcut predictions for the
multipoint correlation functions, which can be tested both
by experiment and by simulations.
The model can be generalized to apply to a broader variety
of situations by including vector forces as well as incorporating
boundary effects.
Most interestingly, we plan to investigate whether the
statistical theory developed here can be extended to provide
new insight into the complex dynamical effects exhibited by
granular materials.\cite{sandrefs}

In summary, we have presented and analyzed a statistical
model for force inhomogeneities in stationary bead packs.
The model, which predicts that force inhomogeneities
decay exponentially at large forces for almost all contact
distributions, agrees well with numerical simulations of
sphere packings as well as experiment.\cite{liu95}


\acknowledgements
We thank S. Nagel for many useful conversations as well as
collaboration
and A. Pavlovitch for discussions and for providing us with
Ref.~\cite{Pav94} prior to publication.
O.N. acknowledges support by the Society of Fellows at
Harvard University and the Institute for Theoretical Physics
at Santa Barbara (NSF Grant No. PHY89-04035).
S.N.C. acknowledges support as an MRL visitor at the
University of Chicago.
This work was supported in part by the MRSEC Program of the
National Science Foundation under award numbers DMR-8819860
and DMR-9400379.
\appendix
\section{The uniform \q-distribution}

Here we consider the ``uniform'' $q$-distribution, which is the
simplest $q$-distribution consistent with the
restriction that $\sum_{1=1}^N q_i = 1$.
It is obtained by choosing each of $q_1,q_2,\ldots,q_{N-1}$
independently from a uniform distribution between $0$
and $1$, setting $q_N= 1-\sum_{j=1}^{N-1}q_j$,
and then keeping only those sets where $q_N \geq 0$.
Here we show that
$\eta_u(q) = \frac{1}{N-1}(1-q)^{N-2}$ for this distribution.

For $N=2$, if one chooses $q_1$ between 0 and 1,
then $q_2=1-q_1$ must also be between 0 and 1, so
that $\eta_u(q)=1$.
When $N=3$, configuration will be retained
only if $q_1 + q_2 + q_3 \leq 1$.
Therefore, the probability of obtaining a value of q
is given by:
\be
\eta_u(q) = M \int_0^{1} dq_1 \int_0^{1} dq_2
 \delta(1-q_1-q_2-q) = M\int_0^{1-q}dq_1 = M(1-q) \ ,
\ee
where $M$ is a normalization constant.
Since $\int_0^1 dq \eta(q) = 1$, one immediately
finds $M=2=(N-1)$.

For general $N$, $\eta_u(q)$ can be written:
\be
\eta_u(q) = V(q)/\int_0^1V(q) dq \ ,
\ee
where
\be
V(q) = \int_0^1 dq_2 \int_0^{1-q_2} dq_3 \ldots
\int_0^{1-\sum_{i=1}^{N-1} q_i} dq_n
 \delta(1 - q - \sum_{i=2}^N q_i) \ .
\ee
Using the identity
\be
\int_0^{1-\sum_{j=1}^{n-k}q_m} (1 -
 \sum_{m=1}^{n-k+1}q_m)^{k-1} dq_{n-k+1}
 = \frac{1}{k}(1-\sum_{m=1}^{n-k} q_m)^k \ ,
\ee
one can show that
\be
\eta_N(q) = (N-1)(1-q)^{N-2}\ .
\ee



%

%

\begin{references}
\bibitem{tunefootnote} In our force chain context, criticality is
		dependent on tuning the distribution of $q$'s so that
		only the values $q=0$ and $q=1$ are possible.
\bibitem{vote}  The voter model is a system of voters with opinions
                (either 1 or 0).
                At independent exponential times, an individual
                reassesses his view by choosing a friend at random
                with certain probabilities and then adopting the
                position of the friend.
\bibitem{sandpile} The use of the term ``sandpile'' in this context
                is rather unfortunate, since we are exploiting a
                mathematical equivalence and do not suggest that
                the dynamical properties of models of SOC are
                directly related to the static properties of
                bead packs.
\bibitem{normnote} The claim that $P_D(v)$ becomes independent
		of $D$ as $D \rightarrow \infty$ is supported by the
		results of numerical simulations as well as by the
		fact that we obtain a consistent analytic solution
		for the distribution $P_{D \rightarrow \infty} (v)$.
\bibitem{newnote}  Note that in the $q_{0,1}$ limit, Eq.~(\ref{Preceqn})
		has a singular solution, $\tilde{P}(s) = 1$, that
		disagrees with the exact results.
		This is because (as can be seen by doing a similar
		mean field analysis in the $q_{0,1}$ limit in terms
		of $Q_D(w)$) the terms proportional to $1/D$
		dropped in going from Eq.~(\ref{Ptildeeqn}) to
		Eq.~(\ref{Preceqn}) are important for the
		critical case.
		This is reasonable physically; for a
		critical system, although the injection process
		becomes less and less important in terms of the
		normalized variables as $D \rightarrow \infty$,
		the ``relaxation time'' (the number of levels one
		must propagate downwards before a perturbation
		$\delta P(v)$ decays away) diverges, so that
		$P(v)$ never settles to an equilibrium form,
		and the injection cannot be ignored.
		No such complications are expected away from
		criticality, where the relaxation time should
		be finite.
\bibitem{prodnote} Some $\rho(q_1,\ldots,q_N)$ that are not of
		product form induce $\eta(q)$'s which have
		$\delta$-functions at $q=0$ and not at $q=1$
		(but not vice versa).
\bibitem{alphanote} The $q$'s are related to the $\alpha$'s via:
		$q_1=\alpha_2$,
		$q_2 = \frac{1}{2}(1-\sqrt{3}\alpha_1-\alpha_2)$,
		$q_3 = \frac{1}{2}(1+\sqrt{3}\alpha_1-\alpha_2)$.
\bibitem{threenote} For beads with more than three contacts, say
		$f_1$, $f_2$, $f_3$, and $f_4$, only the three
		strongest contacts were picked and renormalized to
		derive the $q$'s, namely $q_1 = f_1/(f_1+f_2+f_3),$ etc.
\end{references}
\end{document}